\documentclass[acmsmall,nonacm]{acmart}
\renewcommand\footnotetextcopyrightpermission[1]{}
\AtBeginDocument{%
  \providecommand\BibTeX{{%
    \normalfont B\kern-0.5em{\scshape i\kern-0.25em b}\kern-0.8em\TeX}}}

\acmYear{2022}

\acmJournal{JACM}
\acmVolume{37}
\acmNumber{4}
\acmArticle{111}
\acmMonth{8}

\usepackage{subfig}
\usepackage{graphicx}
\usepackage{longtable}

\begin{document}

\title{Random (Un)rounding : Vulnerabilities in Discrete Attribute Disclosure in the 2021 Canadian Census}

\author{Chris West}
\email{c6west@uwaterloo.ca}
\author{Vecna}
\email{vvecna@uwaterloo.ca}
\author{Raiyan Chowdhury}
\email{rh2chowd@uwaterloo.ca}
\affiliation{%
  \institution{University of Waterloo}
  \city{Waterloo}
  \state{Ontario}
  \country{Canada}
}



\begin{abstract}
  The 2021 Canadian census is notable for using a special form of privacy, random rounding, which independently and probabilistically rounds discrete numerical attribute values. In this work, we explore how hierarchical summative correlation between discrete variables allows for both probabilistic and exact solutions to attribute values in the 2021 Canadian Census disclosure. We demonstrate that, in some cases, it is possible to "unround" and extract the original private values before rounding, both in the presence and absence of provided population invariants. Using these methods, we expose the exact value of 624 previously private attributes in the 2021 Canadian census disclosure. We also infer the potential values of more than 1000 private attributes with a high probability of correctness. Finally, we propose how a simple solution based on unbounded discrete noise can effectively negate exact unrounding while maintaining high utility in the final product.
\end{abstract}


\maketitle

\section{Introduction}
Collecting a census of population is an important undertaking.
Uses of census data include planning services such as schools and childcare, defining voting districts, and modeling diseases such as COVID-19 \cite{census-2021-privacy}. 
However, the collection and publication of data should not be taken lightly.
Even when only aggregate statistics are published, individual personal information can sometimes be recovered \cite{Garfinkel2018UnderstandingDR}. 
Thus, when releasing such statistics, one must use some mechanism to protect the privacy of individuals represented within the dataset.
This is not only a matter of ethics; it is also a matter of policy, as the release of data may be subject to various regulations.
Statistics Canada, which collects and disseminates the Canadian census data, has legal obligations from the Statistics Act \cite{statistics-act} and the Privacy Act \cite{privacy-act} to protect the privacy of data subjects. Similarly, the United States under Title 13 protects against the disclosure of personally-identifiable census data with both severe monetary penalties and potential prison time for a failure to do so \cite{title13}.

The United States Census Bureau used a form of differential privacy to protect the published results from its 2020 census \cite{abowd20222020}. 
Differential privacy provides a formal, mathematically provable privacy guarantee.
The main privacy protection used in the most recent Canadian census is random rounding \cite{census-2021-privacy}, where counts are probabilistically rounded either up or down to a multiple of 5.
Counts closer to the next multiple of 5 have a higher probability of rounding up, and counts closer to the previous multiple of 5 have a higher probability of rounding down. 
This process is formally defined in \hyperref[random-rounding]{Section 2.1}.

Random rounding aims to provide 1) utility by providing values within a small range of the true attribute value, as well as 2) privacy by hiding the true value within this range. However, as we will show, two key choices in implementation for the 2021 Canadian census disclosure expose significant vulnerabilities in the random rounding process. These choices are 1) treating correlated attribute values as independent in the context of random rounding, as well as 2) publishing population-wide invariants which significantly narrow the scope of enumeration.

Using publicly available Canadian census data \cite{cancensus}, we identify the following contributions from our work:
\begin{enumerate}
    \item We show that hundreds of exact attribute values can be extracted from correlations in rounded data in the presence of population invariants.
    \item We hypothesize and test a similar method that can be used to extract exact attribute values from rounded data that does not require population invariants.
    \item We demonstrate a way to extract probabilistic information about randomly rounded attributes, with and without invariants, and use this method to predict the true value of more than 1000 attributes with high confidence.
    \item We propose a simple alternative to random rounding which both increases net utility while mitigating the potential for exact inference.
\end{enumerate}

Although we do not expose individual-level personal information, the inference of randomly-rounded attributes suggests a weakness in the mechanism and the potential for further exploitation. It is our hope that this work motivates the adoption of more rigorous privacy protections in future census collections and distributions.

\section{Background}

\subsection{Random Rounding}
\label{random-rounding}
Canada's 2021 Census of Population uses two main methods to protect the privacy of its subjects: suppression of certain results in areas with fewer than 40 people (or 100 in some cases) and random rounding of counts \cite{census-2021-privacy}. 
Statistics Canada kindly clarified the random rounding algorithm for us:
All counts are rounded to \textit{a} nearby multiple of 5. Note that this is not to the nearest multiple, but instead follows a probabilistic pattern.
The random rounding algorithm $rround$ on some non-negative integer $x$ can be succinctly expressed as follows:
\begin{definition} $rround$.
\label{rround}
\begin{equation}
rround(x) = 
	\begin{cases}
		x - x \bmod 5 & \text{with probability} \indent 1 - \dfrac{x \bmod 5}{5} \\
		x - x \bmod 5 + 5 & \text{with probability} \indent \dfrac{x \bmod 5}{5}
	\end{cases}
\end{equation}
\end{definition}

\label{appendix-a}
\begin{table}[ht]
\centering
\begin{tabular}{|c|c|c|c|}
    \hline
    \textbf{x} & \textbf{P[$rround(x) = 10$]} & \textbf{P[$rround(x) = 15$]} & \textbf{P[$rround(x) = 20$]} \\
    \hline
    10 & 1 & 0 & 0 \\
    \hline
    11 & 4/5 & 1/5 & 0 \\
    \hline
    12 & 3/5 & 2/5 & 0\\
    \hline
    13 & 2/5 & 3/5 & 0 \\
    \hline
    14 & 1/5 & 4/5 & 0 \\
    \hline
    15 & 0 & 1 & 0 \\
    \hline
    16 & 0 & 4/5 & 1/5 \\
    \hline
    17 & 0 & 3/5 & 2/5 \\
    \hline
    18 & 0 & 2/5 & 3/5 \\
    \hline
    19 & 0 & 1/5 & 4/5 \\
    \hline
    20 & 0 & 0 & 1 \\
    \hline
\end{tabular}
\caption{Outcome probabilities for random rounding on the integers 10-20.}
\label{table-rround-probabilities}
\end{table}

Random rounding gives some form of plausible deniability. For instance, a published value of 15 implies that the true number lies anywhere in the range [11,19], with a higher probability density around the center of 15. This also gives obvious utility; we can say with certainty that the published value is no more than 4 away from the true value in any direction. This may be a large amount for small census subdivisions, but at the town, city, or province level, it can become quite a strong statement. 

However, an important aspect to note is that all attribute values are rounded \textit{individually}, even those which are correlated. For example, if a region publishes the rounded counts:

\begin{itemize}
    \item A1: Those who speak only one of English or French
    \item A2: Those who speak only English
    \item A3: Those who speak only French
\end{itemize}

The published values will be rounded independently, despite the fact that $A1_{True} = A2_{True} + A3_{True}$. We will soon demonstrate that this is a significant design flaw.

As a side note, we have chosen to look at attribute values only greater than 10. This is because random rounding is not as well defined in the range [0-9]. Our preliminary experiments showed inconsistency in random rounding for integers less than 10, and some documentation suggests that a mod-10 rounding scheme is used below 10 (although Statistics Canada was not clear on this point). For simplicity, we stick to larger integers that are well-understood and well-defined. Although this limits the scope of experiments, it is more than sufficient for demonstrating vulnerabilities.

\subsection{Invariants}

An interesting choice for the 2021 Canadian census disclosure was to retain net population \textit{invariants}. Invariants are exact values that are not subject to the randomization or noising process and are retained in the final census disclosure. For example, the city of Thunder Bay has a listed population of 108843 (rather than 108840 or 108845, as would be given by random rounding). Note that the United States 2020 census also chose to include certain invariants, such as state-level population and block-level housing unit counts \cite{uscensus}. There can be a number of reasons to include invariants in a census disclosure. They provide obvious utility, and it may be necessary to have an exact value for policy or legal reasons. In the case of the US census, a major reason for having a state-level population invariant is to provide concrete figures for reapportionment of congressional seats \cite{congsets}. However, we will soon show that the inclusion of even the single net population invariant can have significant consequences for otherwise protected randomly-rounded attributes.

\subsection{SAT and SMT Solvers}

SAT and SMT solvers are a general class of algorithms that are designed to solve satisfiability problems. The most common of these problems is that of the Boolean Satisfiability Problem (SAT), which seeks to find candidate assignments of Boolean variables to satisfy logical statements. Satisfiability Modulo Theories (SMT) solvers generalize the Boolean Satisfiability Problem and thus can solve complex integer programming problems subject to well-defined constraints. 

SAT/SMT solvers have previously been used to great success for database reconstruction attacks. For instance, SAT solvers were used for generating candidate solutions to an example census region as a proof of concept for demonstrating the importance of proper privacy measures in census disclosures  \cite{sat}. Internal experiments on 2010 US census and redistricting data have shown that large-scale linear solvers can come up with candidate solutions, which in some cases may be exact or else share common attributes between them that allow for inference \cite{abowdpres}.

\section{Inference}

\subsection{Exact Inference}

In the context of census attribute disclosure, exact inference means that we can generate a candidate solution to our system of variables before provably showing that this is the only possible solution that satisfies the constraints of the problem. We will demonstrate two different types of exact inference methods; one that is reliant on the presence of invariants, as well as one that has no such requirement.

\subsubsection{Invariant-based Exact Inference}

To see why invariants pose a risk to random rounding attribute inference, it is sufficient to show an example.

We noted earlier that correlated attributes are rounded independently. Let us take the example of the discrete age histograms used at the start of each region block census disclosure. A simplified view of the published age counts is given as:\\

A0: Overall Net Population (Invariant)
\begin{itemize}
    \item A1: Age 0-14
    \subitem - A2 Age 0-4
    \subitem - A3 Age 5-9
    \subitem - A4 Age 10-14
    \item A5: Age 15-64
    \item A6: Age 65+\newline
\end{itemize}

Notice the hierarchical correlations between variables. Ages 0-14, 15-64, and 65+ exactly partition the total population. We also know that the variables A2-A4 exactly partition A1. Therefore, we can design a system of variables such that:\\

$A0_{True} = A1_{True} + A5_{True} + A6_{True}$

$A1_{True} = A2_{True} + A3_{True} + A4_{True}$\\

Seeing as A0 is an invariant, $A0_{True} = A0$. However, A1-A6 are subject to random rounding so they may not take on the same value as the ground truth.

Now, let us consider two edge cases in which exact inference is uniquely possible. For simplicity, we will only consider A0, A1, A5, and A6.

\begin{table}[h]
\begin{tabular}{l|l|l|}
\cline{2-3}
 & Real & Randomly Rounded \\ \hline
\multicolumn{1}{|l|}{A0: Net Population} & 48 (Invariant) & 48 (Invariant) \\ \hline
\multicolumn{1}{|l|}{A1: Age 0-14} & 16 & 20 \\ \hline
\multicolumn{1}{|l|}{A5: Age 15-64} & 16 & 20 \\ \hline
\multicolumn{1}{|l|}{A6: Age 65+} & 16 & 20 \\ \hline
\end{tabular}
\caption{Sample Instance for Random Rounding \#1}
\label{toy-example1}
\end{table}

\begin{table}[h]
\begin{tabular}{l|l|l|}
\cline{2-3}
 & Real & Randomly Rounded \\ \hline
\multicolumn{1}{|l|}{A0: Net Population} & 72 (Invariant) & 72 (Invariant) \\ \hline
\multicolumn{1}{|l|}{A1: Age 0-14} & 24 & 20 \\ \hline
\multicolumn{1}{|l|}{A5: Age 15-64} & 24 & 20 \\ \hline
\multicolumn{1}{|l|}{A6: Age 65+} & 24 & 20 \\ \hline
\end{tabular}
\caption{Sample Instance for Random Rounding \#2}
\label{toy-example2}
\end{table}

\newpage

Following our previous formulation, we know that $A0_{True} = A1_{True} + A5_{True} + A6_{True}$. However, by the bounds of random rounding, we know that:\\

$16 \leq A1_{True} \leq 24$

$16 \leq A5_{True} \leq 24$

$16 \leq A6_{True} \leq 24$\\

As it turns out, when all three attributes lie on the edge-case bound, there is only one viable solution. For instance, in Table ~\ref{toy-example1}, we can see quite naturally that the only possible solution is that $A1=A5=A6=16$ since there is no possible smaller random rounding of 3 variables from 20 which sums to 48. Likewise, for Table ~\ref{toy-example2}, similar logic holds that 24 is the maximum true value for all 3 variables given the published value of 20, and thus $A1=A5=A6=24$ is the only possible solution. 

In fact, to find these kinds of vulnerabilities, it is sufficient to  check if:\\

Let $n =$ Number of correlated sub-attributes, and 

Corr = The list of correlated attributes\\

$\sum_{i=1}^{n} A_{Corr[i]} = Invariant \pm 4n$ \\

Keep in mind that it is not necessary that all attributes are equal as we have shown in the examples so far (this was done only for the sake of simplicity). What is important is that all attributes fall on the same side of maximum or minimum value before rounding to the opposite bound.

\begin{figure}[b]
\centering
\includegraphics[scale=0.6]{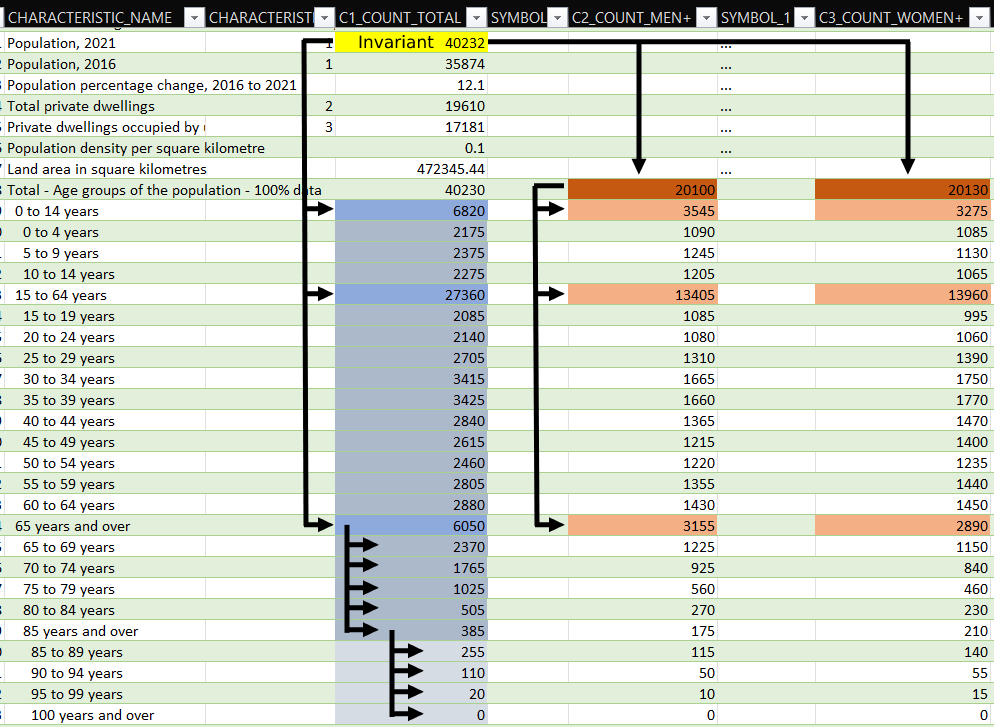}
\caption{Invariant-based Inference, with 1-2 Levels of Potential Compound Inferences Shown}
\label{inf}
\end{figure}

So far we have only demonstrated one possible use of the invariant for age attributes. However, the same logic also holds for sex(+)-based attributes. As shown in Fig \ref{inf}, nearly every attribute in the Canadian census also has the corresponding Men(+) / Women(+) values. (The 2021 Canadian census chose to use a binary division for sex and gender, so non-binary respondents are partitioned into one of these two sex(+) categories~\cite{census-2021-gender}.) These are subject to the same independent random rounding rules as before and are not treated as invariants. Even the net sex(+) population is randomly-rounded.

Solutions generated by exact inference have a worrying potential for further exploitation. Notice that by finding an exact solution to these child attributes, we now open up the possibility for subsequent invariant-based inference. For instance, by inferring the sex(+) invariants in the net population, we open up the possibility of doing the same 3-category age inference attack now not only on the original data but also on the Men(+) and Women(+) attributes for each of the exposes ages. Similarly, in the age example above, by inferring A1 (alongside A5 and A6), we can now attempt a similar attack on A2-A4. That being said, the odds of this type of compound attack succeeding are relatively low due to the unlikely rounding events involved. We did some preliminary testing to check for these compound attacks but did not investigate them thoroughly.

\subsubsection{Invariant-free Exact Inference}

Although invariant-based attacks can be powerful, the only invariant provided in the 2021 Canadian census is the population for each census subdivision (country, provinces, regions, cities, and so on). This means that the scope of our attacks is limited only to those attributes which make up an exact subset of the total population of that subdivision. However, we can show that invariants are not always necessary to do a similar attack to the one shown earlier. Let us demonstrate with an example:

Consider the Celtic Languages category (mother tongue, single response):\\

\begin{itemize}
    \item A581: Celtic Languages
    \subitem A582: Irish
    \subitem A583: Scottish Gaelic
    \subitem A584: Welsh
    \subitem A585: Celtic Languages, n.i.e.\newline
\end{itemize}

As before, we require that the 4 sub-attributes are an exact partition of the parent attribute and that each attribute is independently randomly rounded. Since this is a single response category and n.i.e includes those Celtic languages "not included elsewhere", we can safely assume that this is an exact and mutually exclusive subset. We have included what we believe to be all or at least most of said categories, 20 in total, in Table \ref{tab:cat}.

Unlike in previous examples, we are not given the parent attribute as an invariant; instead, it is also subject to random rounding. Now, let us show two sample instances in tables \ref{toy-example3} and \ref{toy-example4}.

Notice in these two instances that, although we do not have an exact value for the parent attribute, we have a range. For instance, in Table \ref{toy-example3},
 we know that $56 \leq A581_{True} \leq 64$. However, we also know that:\\

$16 \leq A582_{True} \leq 24$

$16 \leq A583_{True} \leq 24$

$16 \leq A584_{True} \leq 24$

$16 \leq A585_{True} \leq 24$\\


\begin{table}[h]
\begin{tabular}{l|l|l|}
\cline{2-3}
 & Real & Randomly Rounded \\ \hline
\multicolumn{1}{|l|}{A581: Celtic Languages} & 64 & 60 \\ \hline
\multicolumn{1}{|l|}{A582: Irish} & 16 & 20 \\ \hline
\multicolumn{1}{|l|}{A583: Scottish Gaelic} & 16 & 20 \\ \hline
\multicolumn{1}{|l|}{A584: Welsh} & 16 & 20 \\ \hline
\multicolumn{1}{|l|}{A585: Celtic Languages, n.i.e.} & 16 & 20 \\ \hline
\end{tabular}
\caption{Sample Instance for Invariant-Free Inference}
\label{toy-example3}
\end{table}

\begin{table}[h]
\begin{tabular}{l|l|l|}
\cline{2-3}
 & Real & Randomly Rounded \\ \hline
\multicolumn{1}{|l|}{A581: Celtic Languages} & 76 & 80 \\ \hline
\multicolumn{1}{|l|}{A582: Irish} & 19 & 15 \\ \hline
\multicolumn{1}{|l|}{A583: Scottish Gaelic} & 19 & 15 \\ \hline
\multicolumn{1}{|l|}{A584: Welsh} & 19 & 15 \\ \hline
\multicolumn{1}{|l|}{A585: Celtic Languages, n.i.e.} & 19 & 15 \\ \hline
\end{tabular}
\caption{Sample Instance for Invariant-Free Inference}
\label{toy-example4}
\end{table}

Since 16*4=64 is the assignment that produces the minimal sum solution, we are certain that the only possible solution is:\\

$A581_{True} = 64$

$A582_{True}, A583_{True}, A584_{True}, A585_{True} = 16$\\

Similar logic follows for the opposite case for the upper bound, but we will not show it for brevity.

In order to check for this vulnerability in a subset of the data, one only has to find attributes with 4 correlated sub-attributes as an exact and mutually exclusive subset, and then check if:\\

Let Corr = The list of correlated attributes\\

$\sum_{i=1}^{4} A_{Corr[i]} = A_{Parent} \pm 20$\\  

Unlike in invariant-based exact inference, note that the number of sub-attributes is constrained. This is because we are invoking edge-case behavior in the summative attribute that only works with certain combinations of both upper and lower bounds. If we consider 2-3 sub-attributes, both the maximum and minimum of their bounds put us in the middle of the parent attribute probability distribution, which has non-deterministic rounding behavior. For instance, 6+6=12, 9+9=18, 6+6+6=18, and 9+9+9=27 all are cases which are not easily exploitable. As it turns out, groups of 4 sub-attributes with a single parent attribute turn out to be the most exciting and exploitable case.

It is theoretically possible to exploit the same vulnerability with 9,14,19 etc. attributes, but the unlikeliness of the rounding becomes vanishingly small with that many attributes. This is because it requires the independent rounding of those many attributes in the exact correct (and unlikely) direction for that attack to work. We will see in \hyperref[sec:results]{Section 4} how even the basic case with 4+1 attributes is statistically very unlikely.

\subsection{Probabilistic Inference}

The final type of inference which we will partially explore is probabilistic inference.

By probabilistic inference, we mean that we now reduce the scope of the problem such that instead of saying "we are certain that attribute $A_i$ takes on value x", we can instead make statements such as "we are 40\% sure that attribute $A_i$ takes on  value x, 30\% that it takes on value x+1, etc."

This is naturally a weaker attack, but it has a much more broad application due to the fact that it does not require  very specific edge-case rounding behavior. As well, depending on the threat model, a 30-50\% certainty in the true value may be sufficient. Not only this, but a probabilistic model can very often rule out other candidate solutions.

Let us consider a small toy example where this may be interesting. If the total population is 87, the posted male(+) population is 35, and the posted female(+) population is 45, we cannot do exact inference. However, notice that there are only 2 possible solutions:\\

M(+) = 38, F(+) = 49 or \\

M(+) = 39, F(+) = 48 \\

This may still be a useful result for an adversary, even in the absence of definitive results. 

For our upcoming experiments, we will explore two different methods for probabilistic inference. One of these methods gives precise statistical information about arbitrary attribute value probability based on SAT solver enumeration but is in general not very practical for finding vulnerabilities in real data. The other is a simple logic-based method similar to those which we have used before. Both can be performed with or without the presence of an invariant.

\subsubsection{Invariant-based Probabilistic Inference}

To determine a precise probability distribution for attribute values, it is possible to employ the use of an SMT solver to enumerate possibilities for solving the system of equations defined by the random rounding constraints. Using our example for age constraints from the previous section, we define something like:\\

$A0 = A1_{True} + A5_{True} + A6_{True}$

$A1-4 \leq A1_{True} \leq A1+4$

$A5-4 \leq A5_{True} \leq A5+4$

$A6-4 \leq A6_{True} \leq A6+4$\\

This is basically the same fundamental system as before so far. However, instead of looking for specific edge cases, we enumerate all possible cases and calculate their relative probabilities. 

First, we generate a potential solution to the system of equations using the Python SMT solver Z3. \cite{z3} Note that the generated solution is only \textit{a} feasible solution; it is not guaranteed to be the only feasible solution and is likely only one of many possible solutions. We must enumerate all other solutions as well. To do this, we add on the current solution as a NOT constraint and rerun the code to generate another solution. This continues until the space of possible solutions is exhausted. Note that the solution space for a particular instance is not easy to express in closed form, and can easily be on the order of hundreds to even tens of thousands or more solutions, depending on the value of the invariant and the number of correlated variables. That being said, the method theoretically composes very well with additional information about the feature space, since this narrows down the search and enumeration space as well.

Once all solutions have been generated, it is not as simple as looking at common features between solutions since not all solutions are equally probable. Instead, we can observe a relative probability distribution of each solution by the odds of the observed random rounding occurring given that solution. For instance, [19 19 19] rounding to [15 15 15] is much less likely than [16 16 16] rounding to [15 15 15]. Since we have the exact odds of random rounding, we can get the exact odds of a possible rounding and then scale this to the total probability of all feasible solutions in a post-processing step. With this, we can rank the most likely solutions as well as their overall probability in the solution space. We will then be able to say something like "The most likely solution is [12, 26, 40] at 4\% probability, and it is 5\% more likely than the next most common solution. There are exactly 412 unique solutions.".

This is clearly an interesting result but is not always particularly informative in and of itself. This is to say that the most likely solution may still only be vanishingly likely due to the size of the solution space. As well, we are more interested in the value or distribution of features rather than the total solution itself since features are more likely to reveal information about a single individual.

To address this, we extend our framework to examine feature values given the probability of their parent solution. Since the probability of the parent solution already takes into account the odds of a given random rounding, we only need to iterate through our solutions and construct the probability histogram for all possible solutions to each feature individually. This is what we will demonstrate on some sample census data in \hyperref[sec:results]{Section 4}.

This method gives precise information about all attributes in the system but is computationally costly. Since the focus of this paper is on looking for specific vulnerabilities in the Canadian 2021 census data, it is also possible to use a simple logic-based method to find non-exact invariant-based inferences. We can simply check for a condition like:\\

Let $n =$ Number of correlated sub-attributes, and 

Corr = The list of correlated attributes\\

$\sum_{i=1}^{n} A_{Corr[i]} = Invariant \pm (4n-1)$ \\

Notice that this is almost identical to the previous condition for exact invariant-based inference, although it is slightly looser. The result of this is that we can infer the value of all but one of the attributes in the system, and thus can say that we are confident about the value of the attribute to a probability of $1 - \frac{1}{n}$ for $n \geq 2$. As we will explore later, this type of attack can still be relatively strong while also being much more likely than the exact-inference equivalent. It is also much less demanding computationally, and so is what we use in practice.

\subsubsection{Invariant-free Probabilistic Inference}

We will also demonstrate that we can do similar methods to the above without the need for an invariant. To show this, let us pretend we are analyzing one of the many attributes which are non-invariant and have both a male(+) and female(+) component. 

We will essentially format the same general system of equations as above, but we no longer have an equality equation to limit the system. Instead, our invariant only has a rough bound: \\

$A0_{Male,True} + A0_{Female,True} >= A0 - 4$

$A0_{Male,True} + A0_{Female,True} <= A0 + 4$

$A0-4 \leq A0_{True} \leq A0+4$

$A0_{Male}-4 \leq A0_{Male,True} \leq A0_{Male}+4$

$A0_{Female}-4 \leq A0_{Female,True} \leq A0_{Female}+4$\\

As before, we can enumerate all possible solutions, rank them by the relative odds of that solution based on random rounding, and then extract individual feature probabilities. There will most likely be many more candidate solutions than in the previous case since the invariant is non-fixed, but unlike the previous invariant-free exact inference technique, we can technically do this on any size of grouped features (albeit with limitations to due computational complexity). We will demonstrate this method on arbitrary features from the census data later on.

As before, we can also do another simple logic-based method to look for high probability but non-exact invariant-free inference. We can simply check for a condition like:\\

$\sum_{i=1}^{3} A_{Corr[i]} = A_{Parent} \pm 15$\\  

If we fulfill this condition, we can say that we are confident about the value of the attribute(s) (including the parent) to a probability of $\frac{3}{4}$. This is because it implies a result such as:\\

$A_{Parent} = [30]$
$A_{Children} = [15, 15, 15]$\\

Possible Solutions = \\

$A_{Parent} = 33, A_{Children} = [11, 11, 11]$

$A_{Parent} = 34, A_{Children} = [12, 11, 11]$

$A_{Parent} = 34, A_{Children} = [11, 12, 11]$

$A_{Parent} = 34, A_{Children} = [11, 11, 12]$\\

Again, this is the primary method we will use to look for invariant-free probabilistic inference vulnerabilities as opposed to the SAT solver method.

\section{Results and Statistical Analysis}
\label{sec:results}

\subsection{Exact Inference Results and Analysis}

Please see Tables \ref{tab:agee} and \ref{tab:sexdis} for results on invariant-based exact inference. Altogether, we found 285 sex(+) exact disclosures and 18 age exact solutions. Since our age attack reveals exactly 3 attributes and the sex(+) attack reveals two, we reveal exactly 624 attributes from the census disclosure using invariant-based exact inference.
We found no examples of invariant-free exact inference.

To analyze the frequency of our attacks succeeding, we did some simple probabilistic calculations based on random rounding probability and validated these across the empirical results. 

\subsubsection{Analysis of Invariant-based Exact Inference}

In order for this attack to succeed, it must be the case that all X attributes fall on the border of the same bound (upper/lower), but round to opposite sides. To estimate the probability of this happening, we multiply the probability of each attribute being at the bound (1/5, in mod 5), and then the probability of that attribute rounding to the opposite side (1/5 as well given the definition of random rounding). We must do this for all attributes in the candidate solution, and then multiply by two to consider both the upper and lower case. \\

In the sex(+) case with only two rounded attributes, this works out to:

$2*(1/5)^4$ = 0.32\% of instances, or 1-in-313.\\

And in the age case with 3 attributes, it works out to:

$2*(1/5)^6$ = 0.013\% of instances, or 1-in-7813.\\

Remember that we are only considering regions where attributes are known to be above 10 since the behavior of sub-10 random rounding is not well defined for our attack. There are 61029 query-able census regions in total per our count, but only 61010 regions where we can apply our attack on sex(+), and 59625 regions where we can apply our attack on age. Given our numbers, this predicts that we will find around 195 exact solutions to sex(+) and 8 exact solutions to age from invariance-based exact inference. However, we discovered 285 exact solutions for sex(+) and 18 exact solutions for age respectively. This is within the relative range of our predictions, but the small discrepancy is worthy of future investigation.

\subsubsection{Invariant-Free Exact Inference}

We do a similar analysis to the above. However, we are now dealing with 5 attributes instead of 3. For the four sub-attributes, we require that they all take on a certain value at the bound (1/5, due to mod 5), and then round to the opposite bound (1/5 by random rounding). Next, the parent attribute must then round to the opposite bound (1/5). Finally, we account for both cases of the upper and lower-bound solutions, so we multiply by two. This gives a probability of:\newline

$2*(1/5)^9 = 1.02x10^{-4}$\% of instances, or 1-in-976562.\\

This seems inordinately small, but it is important to consider that this works for all groups of 4+1 attributes without the need for an invariant. We identify at least 20 such groups per census subdivision. After filtering out those groups which have some attributes possibly below 10, we still count 83898 total, which predicts around 0.0859 cases of exact inference in the full census disclosure from exact invariant-free inference. That makes the probability of this type of attack succeeding statistically unlikely but not altogether impossible. If we had an order of magnitude more municipalities or features, we might expect to find an instance of this attack being successful.

\subsection{Probabilistic Inference}

\subsubsection{Invariant-Based Results}

 To demonstrate the SAT solver enumeration method, we can plot some histograms of numerical features from a sample census town (Colville Lake, in this case). To be interesting, we enumerate not only the three outer age categories but also the 3 subset age categories of age 0-14 (by enforcing their random rounding constraints as well as the constraint that they round to the age 0-14 solution). In this specific town, we can be 90\% sure that there are either 27, 28, or 29 people aged 0-14. 

\begin{figure}[H]
\centering
\subfloat[Age 0-14]{\includegraphics[width=4cm]{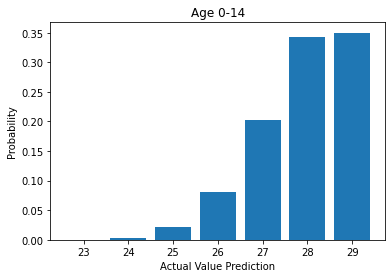}}\hfil
\subfloat[Age 0-4]{\includegraphics[width=4cm]{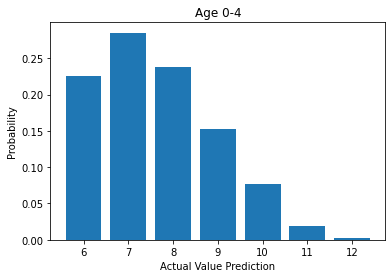}}\hfil 
\subfloat[Age 5-9]{\includegraphics[width=4cm]{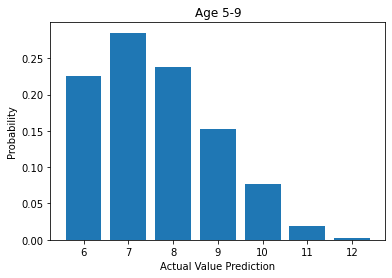}} 

\subfloat[Age 10-14]{\includegraphics[width=4cm]{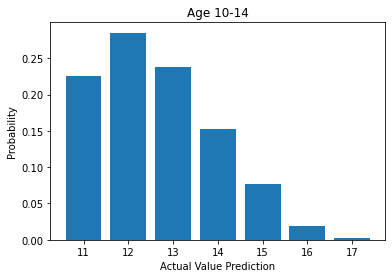}}\hfil   
\subfloat[Age 15-64]{\includegraphics[width=4cm]{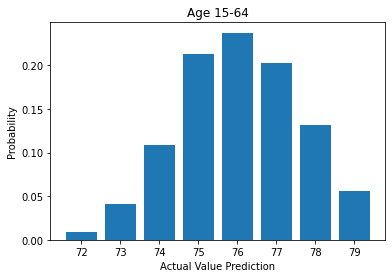}}\hfil
\subfloat[Age 65+]{\includegraphics[width=4cm]{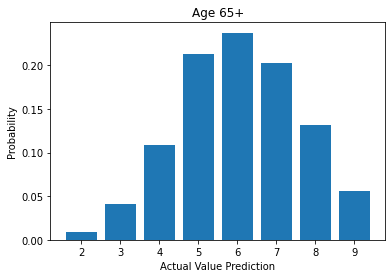}}
\caption{Feature Probability Histograms}\label{figure}
\end{figure}

Next, we also check the census data for the condition we defined earlier. We will define a strong probabilistic result as being at least $66\%$ confident in the value of any given attribute. We chose this value because it works well analytically in our method and gives reasonable confidence in our deductions. By this definition, we examine the age attribute and find 83 examples of strong probabilistic inference in the 2021 Canadian census disclosure. See Table \ref{tab:tabstrongage} for this result.

It's unsurprising that this result is much more common than exact inference. In the age example, we are now allowed 3 different assignments of our three variables that still fulfill the condition, such as: \\

[11 11 12], [11 12 11], [11 11 12] \\

As well, random rounding is more likely to round to the other bound for the odd number out since it is closer to the opposite bound (2/5 odds rather than 1/5). Taking into account both directions of rounding (x2) and all 3 assignments (x3), we find the odds of this invariant-based strong probabilistic attack working on age as: \\

$6 * (1/5)^5 * (2/5)  = 7.68x10^{-2}$\% of instances, or 1-in-1308.\\

This is 6 times more likely than invariant-based exact inference attacks on age. (We will not examine sex(+) for this part, since it can only reach $\frac{1}{2} $ confidence with two sub-attributes for the non-exact case.)\\

This predicts 48 instances of invariant-based strong probabilistic inference on age in the census dataset. Empirically, we find 83. Since each of these enumerates 3 sub-attributes, this is a total of 249 values inferred with strong probability.

As a fun side note, the town of our university, Waterloo, is one of the regions exposed in this attack. We know that:\\

$P[A_{Age 0-14} = 17,644] = \frac{2}{3},$

$P[A_{Age 15-64} = 85,084] = \frac{2}{3},$

$P[A_{Age 65+} = 18,709] = \frac{2}{3}$\\

\subsubsection{Invariant-Free Probabilistic Inference}

 To demonstrate the SAT solver enumeration method, let us explore an arbitrary feature from the data; the number of married spouses or common-law partners in the town of Watson Lake. By using our exhaustive method, we are able to generate a feature probability density without the need for an invariant:

\begin{figure}[H]
\centering
\subfloat[Net Attribute]{\includegraphics[width=4cm]{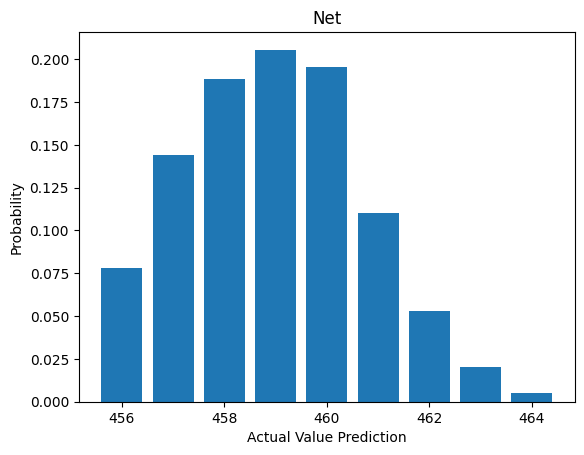}}\hfil
\subfloat[Men(+)]{\includegraphics[width=4cm]{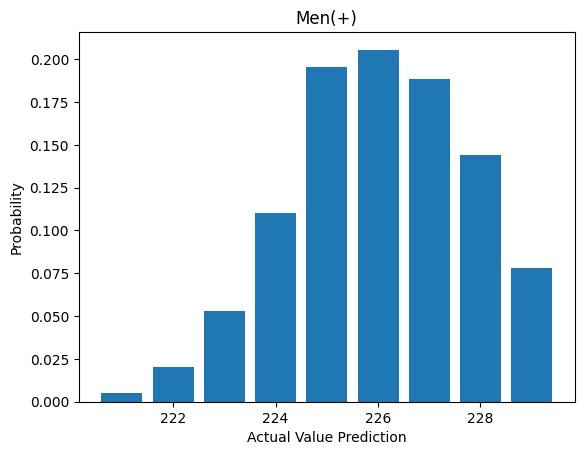}}\hfil 
\subfloat[Women(+)]{\includegraphics[width=4cm]{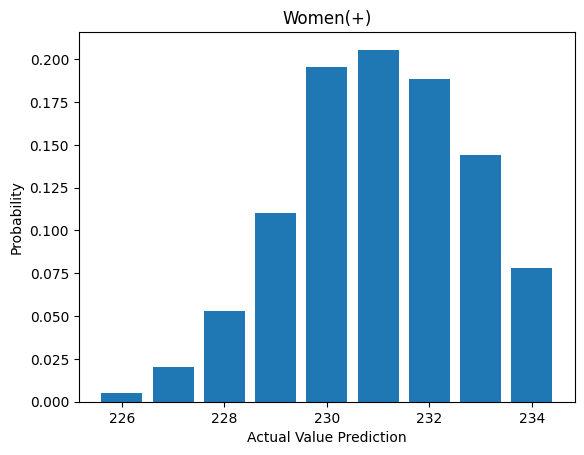}} 
\end{figure}

Again, this is an interesting result and proof of concept but proves to be computationally expensive (and non-revealing for the majority of cases). For this reason, we will mostly explore the census data with the logic condition-based method we defined earlier for invariant-free probabilistic inference. 

Using the previous definition of a strong probabilistic result, we now explore mutually exclusive attribute categories with only 3 sub-attributes. We count 42 of these categories, as shown in Table \ref{tab:probsatts}. We find 216 cases of strong probabilistic inference (with probability P$\geq$0.75 in this case), which reveals 864 individual values with high probability (3 children and 1 parent each). We show these results in Table \ref{tab:strpronon}.

Again, it's unsurprising that this result is much more common than invariant-free exact inference. As explained before, we are allowed multiple assignments of variables that still fulfill the condition for the attack, and the overall random rounding probability is higher.

We first consider the case where the 3 attributes lie on the bound and round with 1/5 probability, such that the summative parent attribute is not bounded and rounds to the opposite bound with 2/5 probability. The second case is when one of the three attributes is one away from the bound as we have shown before. There are two directions of the bound and 3 possibilities per case. As well, the overall probability is the same as before because the subattribute rounds with 2/5 probability whereas the parent attribute now rounds with 1/5 probability.

We find the odds of this strong probabilistic attack working as: \\

$(2 * (1/5)^6 * (2/5)) + (6 * (1/5)^6 * (2/5)) = 2.05x10^{-2}$\% of instances, or 1-in-4882.\\

Since there are 918192 sets of attribute categories that are susceptible to this attack in the census disclosure this predicts 188 instances of invariant-free strong probabilistic inference on age in the census dataset. Empirically, we find 216, which is relatively close to the expected value.

\section{Suggested Solution}

There are a number of potential solutions to the issues highlighted in this paper. 

One such solution is increasing the random rounding bound. For instance, increasing the rounding range would make extracting solutions, both probabilistically and exactly, numerically intractable and vanishingly unlikely. However, the utility of this solution is questionable. By using a range larger than 5, say 7, you now have non-intuitive bounds. 18 could round to 14 or 21, while 24 could round to 21 or 28. One of the major strengths of standard random rounding is interpretability since multiples of 5 are much easier to visualize and analyze than multiples of 7. As well, 15 could round all the way to 21, which is quite a long distance from the ground truth. By using a more intuitive bound like 10, the results once again become easily interpretable, but the utility is much less since any given value can be as much as 9 away in some cases.

Instead, we propose adding unbounded noise to the truth value. By doing this, it suddenly becomes impossible to say with certainty the range of values any specific attribute can take on, so the upper/lower bound exact inference attacks are no longer possible. It also makes SAT solver enumeration attacks intractable since it is generally no longer possible to enumerate all combinations of variables in an unbounded range.

For our solution, we specifically propose using the discrete Laplace distribution with the parameter t=1.45.\\

$\mathbb{P}_{X \leftarrow Lap_{\mathbb{Z}}(t)}[X = x]= \frac{e^{1/t}-1}{e^{1/t}+1}e^{-|x|/t}$ for all $x \in \mathbb{Z} $\\

We choose the discrete Laplace partly because it is peaky and easily sampled. It also has appealing behavior under differential privacy \cite{gautam}, which allows for future extensions and experimentation into quantifiable privacy measures for the Canadian census.

We choose this specific t parameter value because the resulting distribution roughly mimics the random rounding PDF in terms of containing the majority of its probability density within 4 of the ground truth (remember that the greatest distance possible in mod-5 random rounding is 4). Specifically, this discrete Laplace with t=1.45 contains more than 95\% of the expected sample utility within 4 of the true value, which roughly coincides with passing a hypothesis test of p=0.05 for falling within the boundary of random rounding.

\begin{figure}[H]
\centering
\subfloat[Random Rounding Signed PDF]{\includegraphics[width=6cm]{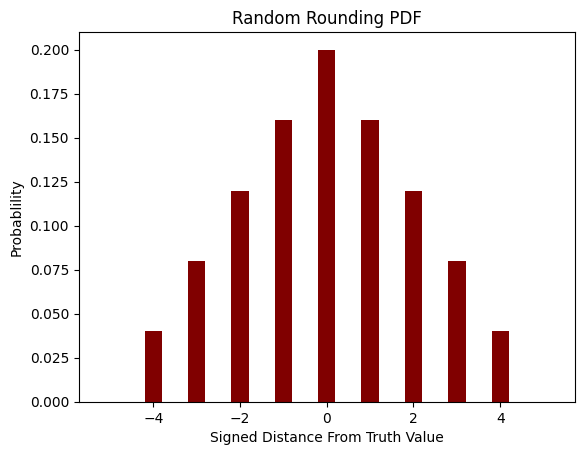}}\hfil
\subfloat[Discrete Laplace Signed PDF]{\includegraphics[width=6cm]{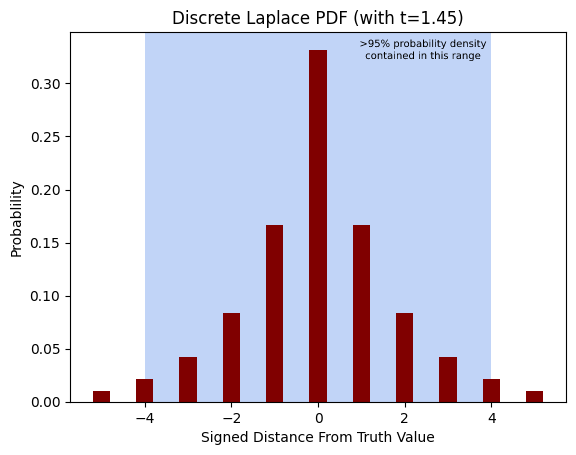}} 
\caption{Signed Distance PDFs}\label{figure}
\end{figure}

Although the privacy benefit is obvious, it is not immediately clear that the utility is comparable to random rounding. Interestingly, we can show that the utility of our approach is actually \textit{superior} to regular random rounding in the average case. That is to say that, on average, our disclosed attribute values are closer to the true value than random rounding, while also having better privacy guarantees against exact inference. To show this, we will investigate the average (non-sub-10) unsigned distance from the true value by enumerating all possible cases as shown in Table \ref{table6}. Notice that sample probability is equal for modular cases since the individual attributes of the exact cover are sampled independently and generally cover the mod 5 space with equal probability.

\begin{table}[H]
\begin{tabular}{llll|l|}
\hline
\multicolumn{1}{|l|}{X mod 5} & \multicolumn{1}{l|}{Sample Prob.} & \multicolumn{1}{l|}{Distance Prob.} & Sum Distance & Weighted Sum Distance \\ \hline
\multicolumn{1}{|l|}{0} & \multicolumn{1}{l|}{0.2} & \multicolumn{1}{l|}{1.0 * 0} & 0 & 0.2 * 0 = 0 \\ \hline
\multicolumn{1}{|l|}{1} & \multicolumn{1}{l|}{0.2} & \multicolumn{1}{l|}{(0.8 * |-1|) + (0.2 * |4|)} & 1.6 & 0.2 * 1.6 = 0.32 \\ \hline
\multicolumn{1}{|l|}{2} & \multicolumn{1}{l|}{0.2} & \multicolumn{1}{l|}{(0.6 * |-2|) + (0.4 * |3|)} & 2.4 & 0.2 * 2.4 = 0.48 \\ \hline
\multicolumn{1}{|l|}{3} & \multicolumn{1}{l|}{0.2} & \multicolumn{1}{l|}{(0.4 * |-3|) + (0.6 * |2|)} & 2.4 & 0.2 * 2.4 = 0.48 \\ \hline
\multicolumn{1}{|l|}{4} & \multicolumn{1}{l|}{0.2} & \multicolumn{1}{l|}{(0.2 * |-4|) + (0.8 * |1|)} & 1.6 & 0.2 * 1.6 = 0.32 \\ \hline
 &  &  &  & Sum Weighted Distance = 1.6 \\ \cline{5-5} 
\end{tabular}
\caption{Determining Sum Weighted Distance Utility for Random Rounding}\label{table6}
\end{table}

We cannot get an exact closed-form distance for the proposed method because the range of the discrete Laplace is unbounded. However, we can estimate a very precise analytical distance by summing the discrete Laplace over a generous range (-10 to 10, for example). The sampled noise value is essentially guaranteed to be within this range (> 99.9\% of samples are within this range). What we find is that the average weighted distance is only ~1.33, which is approximately 20\% closer than the 1.6 average distance given by random rounding. However, unlike in random rounding, it is no longer possible to enumerate edge cases such that we can say with certainty the actual truth values.

Our proposed method may need additional thought when it comes to the zero case since disclosing a value as below zero is non-intuitive and clearly shows an impossible solution. However, we naively suggest either 1) leaving it as is, since exact inference is still impossible even in this case, or 2) clamping the disclosure at 0, such that all published attribute values are non-negative. Further investigation is warranted here if this is to be explored as a real possibility.

\section{Conclusion and Future Work}

A natural question to ask is if these leaked attributes pose any kind of security or privacy risk. This is an open question. It is the authors' opinion that, since these vulnerabilities allow for the reconstruction of non-invariant attributes, they are worthy of patching. It's also a possibility that the exact values obtained from these attacks could be used for more sophisticated attacks. At the present moment, these findings present a vulnerability in the methodology but not an imminent personal risk.

We also believe that we have not nearly exhausted the possibilities for attacking the random rounding mechanism. There are likely more creative ways to perform both exact and probabilistic inference which have not yet been thought out. As well, we chose to ignore sub-10 random rounding because it behaves slightly differently than typical mod-5 random rounding. In the future, we would like to remove this constraint as it would open up a vast number of possibilities for further exploitation.

 Next, we plan to investigate extending our attacks into revealing more discrete information about the underlying dataset. For instance, using SAT solvers to identify or reconstruct the characteristics of certain individuals in the dataset (like the exact ages of people in a region). This would be a much more devastating attack with direct and dire consequences if this type of information could be extracted from the publicly-available census disclosure. We believe this type of attack might be possible with more sophisticated methods, such as by leveraging the median and mean attributes that are published alongside the counting attributes.

 Overall, we have demonstrated a potential vulnerability in the 2021 Canadian census disclosure, and we encourage further research in this area to ensure confidentiality and trust in the security of all future census disclosures.

\begin{acks}
We thank Prof. Gautam Kamath for editing and help with privacy concepts. Thanks to Candace Trusty from Statistics Canada for clarifying the Canadian census's practices.
\end{acks}

\bibliographystyle{ACM-Reference-Format}
\bibliography{sample-base}

\newpage

\appendix

\section{Appendix}

\subsection{Non-Exact Inference}

\fontsize{7}{9}\selectfont
\begin{longtable}{|l|l|l|l|l|l|l|l|}
\caption{All Discovered Age Exact Inference Disclosures}
\label{tab:agee}\\
\hline
Region / Municipality & Pop (Invariant) & A0-14 (Pub) & A15-64 (Pub) & A65+ (Pub) & A0-14 (Real) & A15-64 (Real) & A65+ (Real) \\ \hline
\endfirsthead
\endhead
35180801 & 2192 & 390 & 1275 & 515 & 394 & 1279 & 519 \\ \hline
35190780 & 3882 & 745 & 2760 & 365 & 749 & 2764 & 369 \\ \hline
35290263 & 863 & 155 & 575 & 145 & 151 & 571 & 141 \\ \hline
35310157 & 552 & 75 & 320 & 145 & 79 & 324 & 149 \\ \hline
Chippewas of Rama First ... & 998 & 170 & 700 & 140 & 166 & 696 & 136 \\ \hline
35560318 & 412 & 30 & 195 & 175 & 34 & 199 & 179 \\ \hline
12090879 & 643 & 90 & 475 & 90 & 86 & 471 & 86 \\ \hline
12100084 & 457 & 55 & 300 & 90 & 59 & 304 & 94 \\ \hline
12110069 & 382 & 50 & 245 & 75 & 54 & 249 & 79 \\ \hline
47060411 & 498 & 115 & 345 & 50 & 111 & 341 & 46 \\ \hline
48062615 & 2268 & 620 & 1440 & 220 & 616 & 1436 & 216 \\ \hline
48070161 & 498 & 80 & 315 & 115 & 76 & 311 & 111 \\ \hline
48112456 & 157 & 20 & 85 & 40 & 24 & 89 & 44 \\ \hline
48110853 & 257 & 25 & 195 & 25 & 29 & 199 & 29 \\ \hline
48150153 & 702 & 90 & 425 & 175 & 94 & 429 & 179 \\ \hline
24230246 & 1272 & 125 & 820 & 315 & 129 & 824 & 319 \\ \hline
24530113 & 573 & 95 & 365 & 125 & 91 & 361 & 121 \\ \hline
24710197 & 548 & 100 & 330 & 130 & 96 & 326 & 126 \\ \hline
\end{longtable}

\fontsize{7}{9}\selectfont
\begin{longtable}{|l|l|l|l|l|l|}
\caption{All Discovered Sex(+) Exact Inference Disclosures}
\label{tab:sexdis}\\
\hline
\textbf{Region / Municipality} & \textbf{Pop (Invariant)} & \textbf{Published Male+} & \textbf{Published Fem+} & \textbf{Real Male+} & \textbf{Real Fem+} \\ \hline
\endfirsthead
\endhead
35010267 & 412 & 200 & 220 & 196 & 216 \\ \hline
35060033 & 398 & 205 & 185 & 209 & 189 \\ \hline
35060864 & 638 & 330 & 300 & 334 & 304 \\ \hline
35061314 & 618 & 310 & 300 & 314 & 304 \\ \hline
35070195 & 597 & 305 & 300 & 301 & 296 \\ \hline
35070324 & 488 & 235 & 245 & 239 & 249 \\ \hline
35090110 & 907 & 410 & 505 & 406 & 501 \\ \hline
35100194 & 502 & 220 & 290 & 216 & 286 \\
\hline
\end{longtable}

\begin{longtable}{ll|lll|lll|}
\caption{All Discovered Age Strong Probabilistic Inference Disclosures}
\label{tab:tabstrongage}\\
\cline{3-8}
 &  & \multicolumn{3}{c|}{\textbf{Published}} & \multicolumn{3}{l|}{\textbf{Strong Probs (P \textgreater{}= 0.66)}} \\ \hline
\endfirsthead
\endhead
\multicolumn{1}{|l|}{\textbf{Region / Municipality}} & \textbf{Pop (Invariant)} & \multicolumn{1}{l|}{\textbf{A0-14}} & \multicolumn{1}{l|}{\textbf{A15-64}} & \textbf{A65+} & \multicolumn{1}{l|}{\textbf{A0-14}} & \multicolumn{1}{l|}{\textbf{A15-64}} & \textbf{A65+} \\ \hline
\multicolumn{1}{|l|}{35010302} & 629 & \multicolumn{1}{l|}{115} & \multicolumn{1}{l|}{340} & 185 & \multicolumn{1}{l|}{111} & \multicolumn{1}{l|}{336} & 181 \\ \hline
\multicolumn{1}{|l|}{35060647} & 381 & \multicolumn{1}{l|}{75} & \multicolumn{1}{l|}{245} & 50 & \multicolumn{1}{l|}{79} & \multicolumn{1}{l|}{249} & 54 \\ \hline
\multicolumn{1}{|l|}{35070179} & 511 & \multicolumn{1}{l|}{65} & \multicolumn{1}{l|}{325} & 110 & \multicolumn{1}{l|}{69} & \multicolumn{1}{l|}{329} & 114 \\ \hline
\multicolumn{1}{|l|}{35120390} & 416 & \multicolumn{1}{l|}{80} & \multicolumn{1}{l|}{250} & 75 & \multicolumn{1}{l|}{84} & \multicolumn{1}{l|}{254} & 79 \\ \hline
\multicolumn{1}{|l|}{35150085} & 559 & \multicolumn{1}{l|}{75} & \multicolumn{1}{l|}{350} & 145 & \multicolumn{1}{l|}{71} & \multicolumn{1}{l|}{346} & 141 \\ \hline
\multicolumn{1}{|l|}{35160132} & 386 & \multicolumn{1}{l|}{35} & \multicolumn{1}{l|}{220} & 120 & \multicolumn{1}{l|}{39} & \multicolumn{1}{l|}{224} & 124 \\ \hline
\multicolumn{1}{|l|}{35180190} & 496 & \multicolumn{1}{l|}{65} & \multicolumn{1}{l|}{385} & 35 & \multicolumn{1}{l|}{69} & \multicolumn{1}{l|}{389} & 39 \\
\hline
\end{longtable}

\begin{longtable}{lll|llll|llll|}
\caption{All Discovered Invariant-Free Strong Probabilistic Inferences}
\label{tab:strpronon}\\
\cline{4-11}
 &  &  & \multicolumn{4}{c|}{\textbf{Published}} & \multicolumn{4}{c|}{\textbf{Strong Probs (P = 0.75 \textgreater{}0.66}} \\ \hline
\endfirsthead
\endhead
\multicolumn{1}{|l|}{\textbf{Region / Municipality}} & \multicolumn{1}{l|}{\textbf{Attributes}} & \textbf{Sex(+)} & \multicolumn{1}{l|}{\textbf{A1}} & \multicolumn{1}{l|}{\textbf{A2}} & \multicolumn{1}{l|}{\textbf{A3}} & \textbf{A4} & \multicolumn{1}{l|}{\textbf{A1}} & \multicolumn{1}{l|}{\textbf{A2}} & \multicolumn{1}{l|}{\textbf{A3}} & \textbf{A4} \\ \hline
\multicolumn{1}{|l|}{35020178} & \multicolumn{1}{l|}{{[}335, 336, 338, 339{]}} & F(+) & \multicolumn{1}{l|}{245} & \multicolumn{1}{l|}{25} & \multicolumn{1}{l|}{95} & 110 & \multicolumn{1}{l|}{241} & \multicolumn{1}{l|}{29} & \multicolumn{1}{l|}{99} & 114 \\ \hline
\multicolumn{1}{|l|}{35060073} & \multicolumn{1}{l|}{{[}335, 336, 338, 339{]}} & F(+) & \multicolumn{1}{l|}{160} & \multicolumn{1}{l|}{35} & \multicolumn{1}{l|}{85} & 55 & \multicolumn{1}{l|}{164} & \multicolumn{1}{l|}{31} & \multicolumn{1}{l|}{81} & 51 \\ \hline
\multicolumn{1}{|l|}{35060588} & \multicolumn{1}{l|}{{[}1452, 1453, 1454, 1455{]}} & All & \multicolumn{1}{l|}{405} & \multicolumn{1}{l|}{220} & \multicolumn{1}{l|}{145} & 25 & \multicolumn{1}{l|}{401} & \multicolumn{1}{l|}{224} & \multicolumn{1}{l|}{149} & 29 \\ \hline
\multicolumn{1}{|l|}{35061091} & \multicolumn{1}{l|}{{[}350, 351, 353, 354{]}} & F(+) & \multicolumn{1}{l|}{225} & \multicolumn{1}{l|}{45} & \multicolumn{1}{l|}{135} & 60 & \multicolumn{1}{l|}{229} & \multicolumn{1}{l|}{41} & \multicolumn{1}{l|}{131} & 56 \\ \hline
\multicolumn{1}{|l|}{35061298} & \multicolumn{1}{l|}{{[}1452, 1453, 1454, 1455{]}} & All & \multicolumn{1}{l|}{310} & \multicolumn{1}{l|}{145} & \multicolumn{1}{l|}{165} & 15 & \multicolumn{1}{l|}{314} & \multicolumn{1}{l|}{141} & \multicolumn{1}{l|}{161} & 11 \\ \hline
\multicolumn{1}{|l|}{35061316} & \multicolumn{1}{l|}{{[}350, 351, 353, 354{]}} & M(+) & \multicolumn{1}{l|}{160} & \multicolumn{1}{l|}{30} & \multicolumn{1}{l|}{115} & 30 & \multicolumn{1}{l|}{164} & \multicolumn{1}{l|}{26} & \multicolumn{1}{l|}{111} & 26 \\
\hline
\end{longtable}

\fontsize{5}{7}\selectfont
\begin{longtable}{lllll}
\caption{Categories Explored for Invariant-Free Exact Inference. Note that we choose to include some attributes that only represent 25\% of the data; these pose little threat because they are already subsampled from the population, but they are interesting targets nonetheless.}
\label{tab:cat}\\
\hline
\multicolumn{1}{|l|}{\textbf{Parent Attribute}} & \multicolumn{1}{l|}{\textbf{Sub \#1}} & \multicolumn{1}{l|}{\textbf{Sub \#2}} & \multicolumn{1}{l|}{\textbf{Sub \#3}} & \multicolumn{1}{l|}{\textbf{Sub \#4}} \\ \hline
\endfirsthead
\endhead
 &  &  &  &  \\
\textbf{Age} &  &  &  &  \\
29: 85 years and over & 85 to 89 years & 90 to 94 years & 95 to 99 years & 100 years and over \\
 &  &  &  &  \\
\textbf{Marital Status} &  &  &  &  \\
61: Living common-law & Living common law ... & Living common law ... & Living common law ... & Living common law ... \\
66: Not married and not ... & Not married and ... & Not married and ... & Not married and ... & Not married and not ... \\
 &  &  &  &  \\
\end{longtable}

\begin{longtable}{llll}
\caption{Categories Explored for Invariant-Free Probabilistic Inference. Note that we choose to include some attributes that only represent 25\% of the data; these pose little threat because they are already subsampled from the population, but they are interesting targets nonetheless.}
\label{tab:probsatts}\\
\hline
\multicolumn{1}{|l|}{\textbf{Parent Attribute}} & \multicolumn{1}{l|}{\textbf{Sub \#1}} & \multicolumn{1}{l|}{\textbf{Sub \#2}} & \multicolumn{1}{l|}{\textbf{Sub \#3}} \\ \hline
\endfirsthead
\endhead
 &  &  &  \\
\textbf{Age} &  &  &  \\
10: 0 to 4 years & 5 to 9 years & 10 to 14 years & 15 to 64 years \\
 &  &  &  \\
\textbf{Family} &  &  &  \\
89: Total - Persons in census families & Married spouses or common-law partn & Parents in one-parent families & Children \\
96: Total - Persons not in census famil & Living alone & Living with other relatives & Living with non-relatives only \\
 &  &  &  \\
\end{longtable}

\end{document}